\documentclass[letterpaper, 12pt]{article}
\usepackage[english]{babel}
\usepackage[utf8x]{inputenc}
\usepackage[T1]{fontenc}
\usepackage{setspace}
\usepackage[a4paper,top=3cm,bottom=2cm,left=2cm,right=2cm,marginparwidth=1.75cm]{geometry}

\usepackage{amsmath}
\usepackage{graphicx}
\usepackage[colorinlistoftodos]{todonotes}
\usepackage[colorlinks=true, allcolors=black]{hyperref}
\usepackage{caption}
\usepackage{subcaption}
\usepackage{sectsty}
\usepackage{float}
\usepackage{titling} 
\usepackage{blindtext}

\usepackage[colorinlistoftodos]{todonotes}
\usepackage{xcolor}
\usepackage{chemformula}
\usepackage[version=4]{mhchem}
\definecolor{darkgreen}{rgb}{0.0, 0.4, 0.0}
\usepackage{cancel}
\usepackage{csquotes}

\usepackage{authblk}

\usepackage[backend=biber,style=numeric,sorting=none,autocite=superscript,maxbibnames=99]{biblatex}
\DeclareFieldFormat{labelnumberwidth}{#1}

\DeclareBibliographyDriver{article}{%
  \printnames{author}
  \setunit{\addcomma\space}%
  \printfield{title}%
  \setunit{\addcomma\space}%
  \printfield{journaltitle}%
  \setunit{\addcomma\space}%
  \printfield{volume}%
  \setunit{\addnbspace}%
  \printfield{number}%
  \setunit{\addcomma\space}%
  \printfield{pages}%
  \setunit{\addcomma\space}%
  \printfield{year}%
  \setunit{\addcomma\space}%
  \printfield{doi}%
  \finentry
}
\let\cite\autocite

\usepackage{textgreek}
\usepackage[labelfont=bf]{caption}
\usepackage{setspace}

\usepackage{pdfpages}
\begin{document}


\title{\textbf{Charge-Transfer Induced Reactivity\\ in \textit{sp} Carbon Atomic Wires:\\ Towards 0–D \textit{sp–sp²} Nanostructures}}

\author[a]{Marco Agozzino}
\author[a]{Eleonora  Moroni}
\author[a,b]{Yifan Zhang}
\author[a]{Valeria Russo}
\author[a]{Carlo Spartaco Casari\thanks{Corresponding author}}

\affil[a]{Department of Energy, Micro and Nanostructured Materials Laboratory – NanoLab, Politecnico di Milano, Via Lambruschini 8, Milano, 20156, Italy}
\affil[b]{School of Engineering, Huzhou Normal University, Huzhou, 313000, China}

\predate{\vspace{-10pt}}  
\date{}  
\postdate{\vspace{-10pt}}  
\maketitle

\begin{abstract}

\noindent Carbon Atomic Wires (CAWs) are finite linear chains of \textit{sp}-hybridized carbon atoms. Here the electrochemical reduction of CAWs in the form of polyynes (i.e. with alternated single-triple bonds) is reported. Upon applying a reducing potential to a solution containing polydispersed hydrogen-capped polyynes, the formation of a black precipitate was observed. Electronic absorption spectroscopy confirmed the irreversible reaction of the carbon chains while excluding degradation or side reactions. Subsequent analyses revealed that the precipitate consisted of amorphous carbon nanoparticles with tunable diameters. This control over particle size is attributed to the modulation of growth kinetics through restricted mass transport toward the solid-liquid interface.\\
Raman spectroscopy showed that the resulting material exhibits an amorphous \textit{sp-sp²} character, with a retained \textit{sp} fraction exceeding 60\%. Smaller nanoparticles displayed reduced disorder within the \textit{sp²} domains and a broader distribution of \textit{sp}-chain lengths preserved in the amorphous matrix. Additional experiments on size-selected polyynes suggest that this synthesis method allows to better preserve the starting chain length in the final structure.\\
Unlike previously reported amorphous \textit{sp-sp²} carbon networks, the nanoparticles produced in this study show remarkable stability under ambient conditions, retaining their \textit{sp} character for times in excess of six months. These findings pave the way for future applications, particularly as further diameter tuning may enable access to the quantum-dot regime.

\end{abstract}

\noindent\textbf{Keywords:} Amorphous Carbon, Nanoparticles, Electrochemistry, Raman Spectroscopy, Polyyne\\
\\
\noindent \textbf{E-mail addresses}\\
\textbf{M. A.:} marco.agozzino@polimi.it, \textbf{E. M.:} eleonora.moroni@mail.polimi.it,\\ \textbf{Y. Z.:} yifan\_zhang@huznu.edu.cn, \textbf{V. R.:} valeria.russo@polimi.it\\
\textbf{C. S. C.:} carlo.casari@polimi.it
\thispagestyle{empty}
\clearpage
\pagenumbering{arabic}
\section{Introduction}
Carbon in the sp hybridization state can be a powerful asset in the design of nanomaterials. Finite \textit{sp}-carbon chains can be regarded as carbon atomic wires (CAWs) and are characterized by extensive \textpi-conjugation. CAWs can be classified as cumulenes or polyynes, depending on whether they contain cumulated double bonds or alternating single and triple bonds \cite{casari_carbyne_2018}. This peculiar electronic structure is key to understanding the unique optical and electronic properties of CAWs. Their strong absorption in the UV-visible range \cite{CATALDO200782,EASTMOND19724601}, exceptional hyperpolarizability \cite{eisler_polyynes_2005,doi:10.1021/acs.jpcc.6b03071}, and large Raman scattering cross-section \cite{RAVAGNAN20061518,doi:10.1021/acs.nanolett.0c02632,marabotti_electron-phonon_2022,MARABOTTI2024118503} fully justify the growing interest in CAWs as prospective materials for biomedical applications \cite{hu_supermultiplexed_2018}, nonlinear optics \cite{MARABOTTI2024118503}, and temperature sensing \cite{doi:10.1021/acsnano.1c03893}. Furthermore, CAWs have been shown to effectively coordinate ions, opening the way to energy storage applications such as lithium-ion batteries and supercapacitors \cite{wang_carbon_2020, park_carbyne_2013,ghosh_cumulenic_2025}.\\

To date, research on \textit{sp}-carbon nanostructures has focused primarily on one- and two-dimensional materials. Significant efforts have been devoted to the study of CAWs as isolated chains stabilized by suitable end groups \cite{doi:10.1021/acscentsci.3c01090}, through polymer encapsulation \cite{peggiani_situ_2020,doi:10.1021/acs.jpcc.5c02960}, or by insertion into nanotubes \cite{shi_confined_2016}. This approach has yielded notable results, such as the fabrication of the first CAW-based field-effect transistor \cite{scaccabarozzi_field-effect_2020} and electrochemical supercapacitor \cite{ghosh_cumulenic_2025}. It should be noted that only CAWs terminated with bulky end groups can be stabilized in the solid state, while other CAWs can only be exploited in solution \cite{casari_carbyne_2018,doi:10.1021/acscentsci.3c01090}.

A notable example of two-dimensional \textit{sp-sp²} hybrid structures that employ \textit{sp} linkers between \textit{sp²} domains to recreate graphene-like order are graphynes and related materials. Despite the remarkable electronic properties of these materials \cite{B922733D,doi:10.1021/acs.jpcc.1c04238}, research progress towards truly 2D atomically precise graphdiyne networks is hindered by the complex synthesis methods required and their intrinsic instability, which prevents exposure to ambient conditions \cite{rabia_structural_2020}.

Only a few reports describe three-dimensional crystalline phases obtained through van der Waals stacking of \textit{sp}-carbon chains, and the size of the ordered domains remains on the order of tens of atoms \cite{pan_carbyne_2015,YANG2022100692}.\\

More extensive progress has been achieved in the investigation of amorphous \textit{sp-sp²} carbon obtained through physical methods such as Supersonic Cluster Beam Deposition from a Pulsed Microplasma Cluster Source (SCBD-PMCS) \cite{PhysRevLett.89.285506}. However, in these studies, the \textit{sp} fraction remains limited to less than 40\%, and the material can only be produced in the form of nanostructured films \cite{ravagnan_influence_2007,RAVAGNAN20061518}. Furthermore, its stability upon exposure to air and elevated temperatures is severely limited, posing a significant obstacle to the study of its intrinsic properties and restricting potential applications \cite{PhysRevLett.89.285506,ravagnan_influence_2007,casari_chemical_2004}. Still, amorphous \textit{sp- sp²} carbon films deposited by SCBD-PMCS under high-vacuum
conditions have been stabilized and tested as electrode materials in ionic liquids, demonstrating the influence of sp-carbon chains on electrochemical double-layer capacitance \cite{Bettini_2016}. The positive influence of the \textit{sp} fraction on the electrical transport in SCBD-PMCS films has also been reported, interpreting the role of the \textit{sp} domains as deep defects state pinning the Fermi level out of midpgap, and thus showing a \textit{self-doping} effect \cite{ravagnan_influence_2007}.\\

Quite recently, CAWs have been proposed as building blocks for novel carbon nanomaterials. In a study by Compagnini and co-workers, carbon nanowalls composed of few-layer graphene were fabricated starting from a solution of polyynic CAWs. A 532 nm nanosecond pulsed laser was employed to ablate a graphite target in an aqueous solution, thereby producing CAWs. During ablation, a pair of electrodes was submerged in the solution, and a potential difference of 30 V was applied. The electrophoresis and subsequent assembly of the CAWs resulted in the formation of a carbon nanowall coating on the cathode \cite{COMPAGNINI20122362}. This technique has since been refined, extending deposition to polyynes synthesized via arc discharge, yielding nanowalls suitable for resistive switching memory devices \cite{RUSSO201754}. More recently, the polymerization of acetylene and the subsequent decomposition of the resulting CAWs was exploited to yield \textit{sp²} carbon nanoparticles with fluorescence properties \cite{JAYSWAL2023337}.\\

The idea of developing a bottom-up approach that relies on an \textit{sp}-hybridized precursor is fascinating. However, in these works, the \textit{sp} character of the reagents is lost, as the final nanomaterial consists purely of \textit{sp²} carbon with a high degree of graphitic order. On the other hand, it has been reported that by controlled reduction of halogenated alkenes \cite{kastner_reductive_1995,kijima_novel_1996,hlavaty_modification_1997} and alkines \cite{KIJIMA19951837}, amorphous \textit{sp-sp²} carbon structures can be synthesized. The possibility of combining these two approaches is the starting point of this work. The main goal is then to study the reactivity of hydrogen-capped polyynes (\ch{HC_nH}, with 8 $\leq$ n $\leq$ 18) assessing weather more moderate reaction conditions can preserve the sp character in the structures that are known to form from previous studies.\\

Indeed, we have achieved amorphous \textit{sp-sp²} carbon nanoparticles with tunable diameter. The effect of process parameters on the morphology and microstructure was investigated by Raman spectroscopy, a well-established tool for the characterization of amorphous carbon materials \cite{PhysRevB.61.14095}. The controlled reduction of size-selected polyynes was further exploited to gain insight into the reaction mechanism.

\section{Experimental Methods}
\subsection*{Polyyne Preparation}
Hydrogen-capped polyynes were synthesized following a modified version of the procedure reported in Ref. \cite{CATALDO20052792}. In a round-bottom flask charged with a magnetic stirring bar, 12 ml ethanol (96\%, Carlo Erba Reagents), 4 ml DI water ($\sigma \leq 0.055$ \textmu S) and 20 ml cyclohexane ($\geq$99.5\%, Sigma-Aldrich) were added. Subsequently, 0.45 g ammonium chloride ($\geq$99.5\%, Sigma-Aldrich), 0.30 g CuCl ($\geq$99\%, Sigma Aldrich), 0.48 g \ch{CuCl2 . 2 H2O} ($\geq$97\%, Sigma-Aldrich) and 0.4 g \ch{CaC2} ($\geq$75\%, Sigma-Aldrich) were added in this order. The reaction mixture was stirred vigorously at room temperature for 30 min. 2 ml of concentrated HCl (37\%, Supelco) were measured and brought with DI water to a total volume of 4ml. After adding the acid solution to the reaction mixture, the organic layer was collected. A second extraction was performed with an additional 10 ml of cyclohexane. The crude product was filtered through silica (0.063-0.200 mm, Labkem) to purify it from byproducts. The refined product was transferred in acetonitrile (MeCN, $\geq$99.9\%, Sigma Aldrich) for electrochemical testing. Size-selected polyynes were separated by reversed-phase High-Performance Liquid Chromatography (RP-HPLC, Shimadzu Prominence UFLC) equipped with a
photodiode array (DAD) UV-Vis spectrophotometer, a fraction collector (FRC-10A), and a C18 column (Phenomenex Luna 5 μm C18(2) 100 Å, LC Column 150 x 10 mm). The gradient method employed is described Ref. \cite{MARABOTTI2022219}.

\subsection*{Electrochemical Synthesis}
The MeCN solution was dried using calcium chloride ($\geq$93.0\%, Sigma-Aldrich) and tetrabutylammonium tetrafluoroborate (TBA-TFB, $\geq$99.0\%, Sigma-Aldrich) was added as a supporting electrolyte. The concentration of \ch{HC8H}, the most abundant polyyne in the mix, was determined from UV-Visible spectra acquired with a Shimadzu UV-1800 spectrophotometer. A molar extinction coefficient $\varepsilon=199550$ L mol$^{-1}$ cm$^{-1}$ was used for the absorption band at 225 nm \cite{CATALDO200782}.\\
Measurements were conducted in a three-electrode cell employing Pt wires (250 \textmu m, 99.99\%, Goodfellow Cambridge Ltd.) as working, counter and reference electrode. Sample volume in the cell was fixed at 3 ml. To study deposition on the electrode surface, gold-coated silicon electrodes were fabricated using a Edwards E306 thermal evaporator. 45 nm Au (99.9\%, Mateck) films were deposited on Si wafers (test grade, Siegert Wafer) adding a 5 nm titanium (97.5\% purity, Sigma-Aldrich) adhesion-promoting layer. Electrochemical measurements were conducted on a Palmsens 4 potentiostat. For each sample, after a preliminary cyclic voltammetry measurement, the potential at the working electrode was kept constant at -1.8 V vs Pt pseudo-reference electrode for 400 s.

\subsection*{Structural and Morphological Characterization}
After electrochemical measurement each sample was allowed to rest and a black precipitate collected on the bottom of each vial. The precipitate was collected and rinsed with fresh acetonitrile to remove leftover polyynes and the electrolyte. The clean precipitate was deposited on a silicon substrate to allow further characterization. Scanning Electron Microscopy (SEM) images were collected using a Zeiss Supra 40 field emission SEM keeping the accelerating voltage at 5 kV to avoid damaging the sample. Raman spectra were collected through a Renishaw InVia MicroRaman spectrophotometer equipped with an \ch{Ar+} laser emitting at 514 nm and a 1800 lines/cm grating. Power on sample was set at 0.05 mW, increasing up to 0.25 mW for low-signal samples.

\section{Results and Discussion}
\begin{figure}[h!]
    \centering
    \includegraphics[width=1\linewidth]{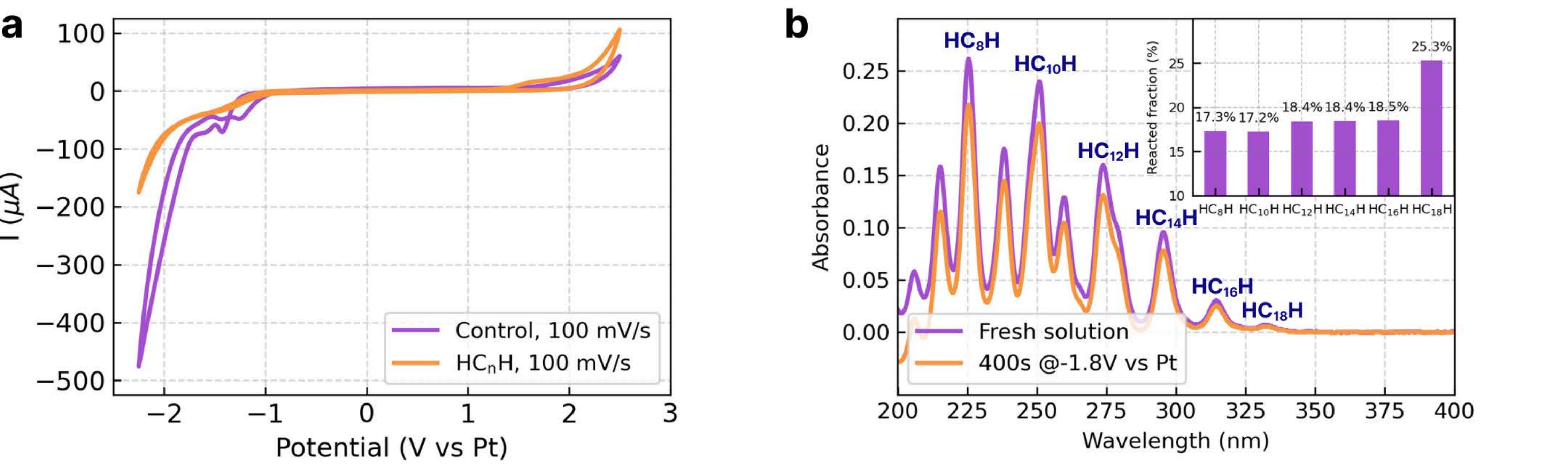}
    \caption{\textbf{Electrochemical synthesis} \textbf{a)} Cyclic voltammogram of polyynes in anhydrous MeCN with 0.1 M TBA-TFB as the supporting electrolyte. The CV of dry MeCN is shown as a reference. \textbf{b)} Electronic absorption spectra of polyynes in MeCN before and after a potentiostatic run. In the inset the polyyne consumption is reported. It is expressed as the fraction of the initial concentration and it has been determined from the intensity variation of the main vibronic peak for each chain length.}
    \label{fig:EC-UV}
\end{figure}

The first evidence of the peculiar electrochemical behavior of polyynes in solution comes from cyclic voltammetry (CV). From Fig. \ref{fig:EC-UV}a, it is evident that the addition of polydispersed H-capped polyynes greatly increases the overpotential required for solvent degradation. The redox peaks observed around -1.5 V vs Pt can be attributed to the oxygen dissolved in MeCN \cite{LI2013328}. This reactivity is quenched by addition of CAWs. Furthermore, when reaching the lower edge of the imposed potential window, a black precipitate can be seen forming around the working electrode. By applying a negative potential (-1.8 V vs Pt) for 400 s, the amount of precipitate further increases. The electronic absorption spectra reported in Fig. \ref{fig:EC-UV}b further confirm the consumption of polyynes upon electrochemical reduction. Quite interestingly, the background usually associated with degradation byproducts is not observed, and the relative decrease in concentration is found to be uniform for all chain lengths (as can be seen from the inset in Fig. \ref{fig:EC-UV}b). The different value obtained for the case of \ch{HC18H} can be ascribed to the uncertainty deriving from its low concentration.

\begin{figure}[h!]
    \centering
    \includegraphics[width=1\linewidth]{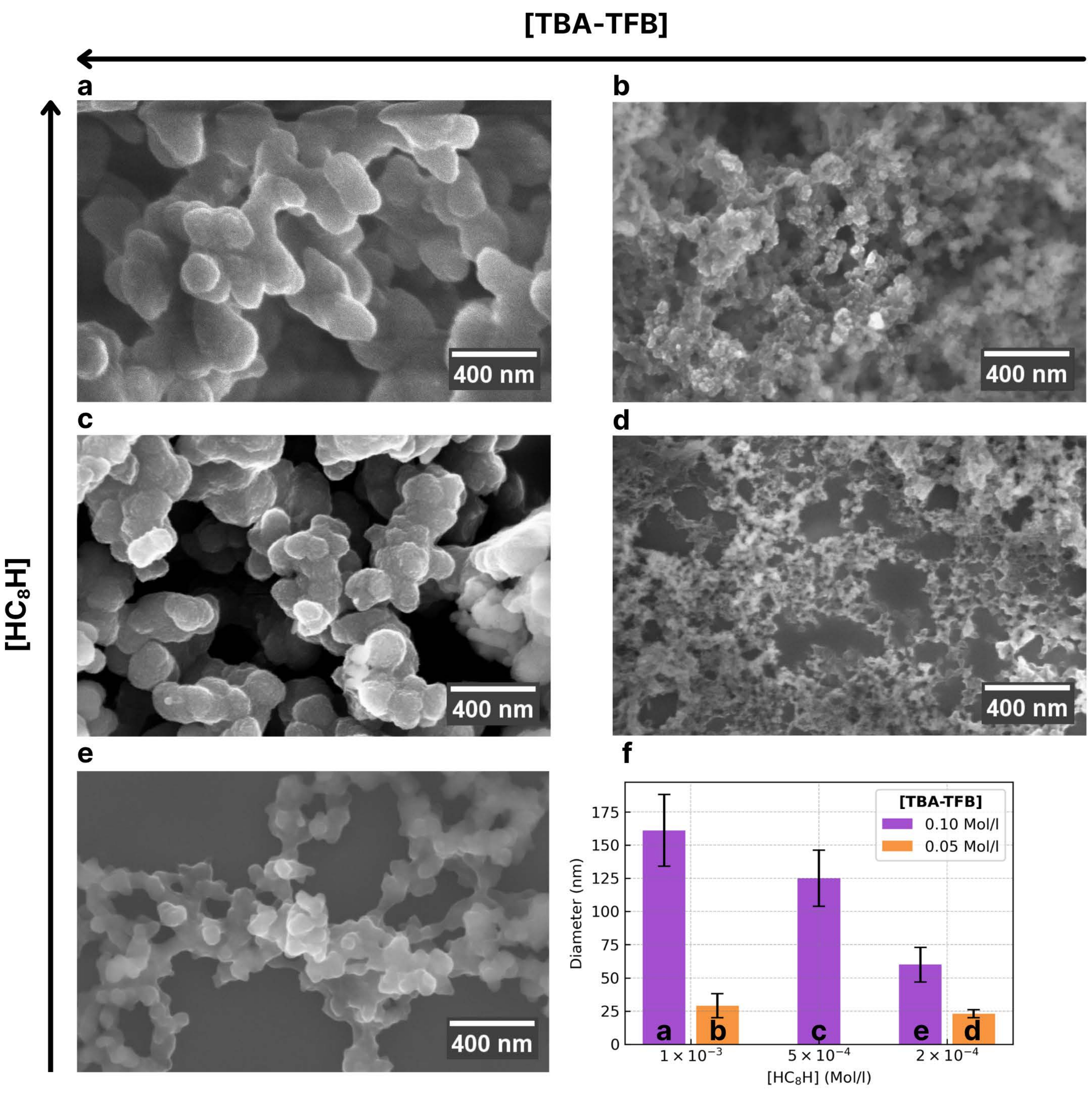}
    \caption{\textbf{Morphological characterization} Scanning Electron Microscopy (SEM) images of nanoparticle aggregates obtained after a 400 s potentiostatic run at $-1.8$ V vs. Pt pseudo-reference from mixed CAWs in ACN solution with a supporting electrolyte concentration of 0.1 M and an \ch{HC8H} concentration of \textbf{a)} $1 \times 10^{-3}$ M, \textbf{c)} $5 \times 10^{-4}$ M, and \textbf{e)} $2 \times 10^{-4}$ M. A second set of experiments was performed with a supporting electrolyte concentration of 0.05 M and an \ch{HC8H} concentration of \textbf{b)} $1 \times 10^{-3}$ M and \textbf{d)} $2 \times 10^{-4}$ M. Higher resolution SEM images of the last two samples can be found in Fig. S1 of the Supplementary Information. The corresponding size distributions are summarized in \textbf{f)}.}
    \label{fig:sem}
\end{figure}

The results obtained by CV are consistent with the formation of a new compound at the electrode surface which, by hindering mass and electron transport, results in sluggish degradation kinetics for the acetonitrile TBA-TFB solution. By collecting the precipitate and performing SEM imaging, it was found that the reduction of polyynes in solution yields a dispersion of nanoparticles. Remarkably, by changing the concentration of polyynes in the solution, it was observed that the size distribution could be tuned. A similar effect was observed upon a reduction in the supporting electrolyte concentration, as evident from the SEM images in Fig. \ref{fig:sem}a-e. The average diameter and standard deviation obtained from the different experimental runs are reported in Fig. \ref{fig:sem}f.\\

A straightforward interpretation of this phenomenon can be found by considering that the size of the nanoparticles generated at the electrode surface is a result of the balance between growth and nucleation. In this instance, the diameter of the nanoparticles is governed by the density of polyynes available to react at the solid–liquid interface. By limiting the precursor concentration in the bulk solution, a shallower gradient is established in the diffusion layer, thus hindering mass transport towards the electrode and slowing down growth.\

The effect of the supporting electrolyte is analogous; by enhancing the solution's conductivity, the reaction rate can be increased as the ohmic drop is reduced, promoting nanoparticle growth.\\

\begin{figure}[h!]
    \centering
    \includegraphics[width=1\linewidth]{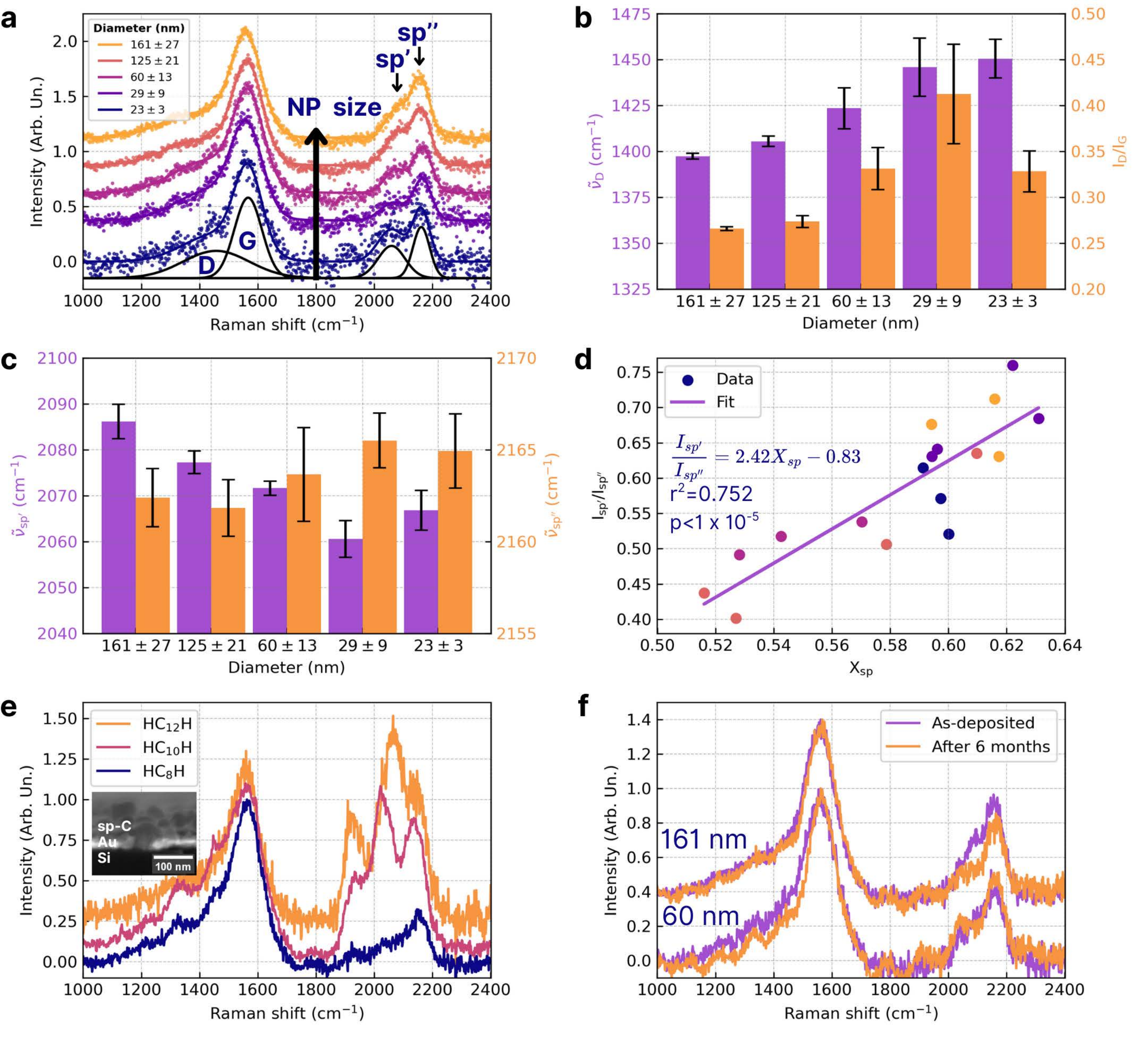}
    \caption{\textbf{Spectroscopic characterization} \textbf{a)} Baseline-corrected Raman spectra of nanoparticle aggregates obtained after a 400 s potentiostatic run at $-1.8$ V vs. Pt pseudo-reference from mixed CAWs in ACN solution. \textbf{b)} D-line Raman shift and intensity ratio between the D and G bands. \textbf{c)} Raman shift of the $sp^\prime$ and $sp^{\prime \prime}$ bands. \textbf{d)} Correlation between the \textit{sp} fraction and intensity ration between the two \textit{sp} components These parameters were retrieved from the spectra in a). \textbf{e)} Raman spectra of a Au-coated Si electrode (SEM image shown in the inset) after a potentiostatic run in an ACN solution containing size-selected CAWs. \textbf{f)} Spectra of the nanoparticles with $d=161 \pm 27$ nm and $d=60 \pm 13$ nm after drying and after six months stored in ambient condition, protected from light exposure. For all Raman spectra the excitation wavelength is 514 nm.}
    \label{fig:raman}
\end{figure}

Raman spectroscopy proved to be a valuable asset in understanding the structure of the nanoparticles. The spectra reported in Fig. \ref{fig:raman}a are characterized by the spectral features of amorphous \textit{sp-sp²} carbon. Both the G and D bands, usually associated with \textit{sp²} carbon, can be discerned, even though the latter is quite weak and broad. Interestingly, a quite intense band is also present in the region between 1900–2200 cm$^{-1}$. This spectral feature is peculiar to materials that incorporate \textit{sp} carbon chains \cite{RAVAGNAN20061518,ravagnan_influence_2007,casari_chemical_2004,Bettini_2016}.
As commonly observed in other disordered \textit{sp-sp²} structures, the vibrational modes give rise to broad bands. Moreover, two components can be discerned in the \textit{sp} band (henceforth referred to as $sp^\prime$ and $sp^{\prime\prime}$, respectively). For this reason, two Gaussians were used to fit the features associated with \textit{sp} carbon. Two Gaussians were also used to fit the D and G bands. This procedure is less informative than the more established approach that describes the D band with a Lorentian and the G band with a Breit-Wigner-Fano (BWF) function \cite{PhysRevB.61.14095}. However, the noise in the spectra renders impractical the implementation of a BWF-Gaussian approach. Indeed, the D band contribution is captured by the asymmetry in the BWF resulting in fitting parameters that are non-physical. A full overview of the spectra processed in this works as well as the results from the fitting procedure can be found in Fig. S2 of the Supplementary Information and in the code shared together with the raw data.

The evolution of the \textit{sp²} component in this material with decreasing nanoparticle dimensions can be traced by considering Fig. \ref{fig:raman}b and more precisely the intensity ratio between the D and G bands ($I_D/I_G$). It is well known that this metric can be reliably related to the degree of disorder in \textit{sp²} domains as it increases with the degree of disordering. This trend is held up to a threshold after which it is reversed, as a further disordering of the structure results in the disgregation of the six-memberded rings that are necessary for the D band signal \cite{PhysRevB.61.14095}.\\ 

It is reasonable to assume that the material studied in this work lies in this last stage of amorphisation and thus an increasing $I_D/I_G$ ratio for smaller nanoparticles can be interpreted as a possible sign of a more ordered structure. While the center of the G band does not vary significantly, a blue-shift of the D band frequency $\tilde{\nu}_D$ is observed. Even if there is no clear consensus on the physical interpretation of the shift in $\tilde{\nu}_D$, it has been proposed that higher frequencies are linked to ordering in highly disordered carbon structures, where the D band also appears at unusually large energies, as is the case in this work \cite{10.1063/1.362745}. This, together with the band's large width suggests a point of caution as it has been pointed out that in highly inhomogeneous structures the interpretation can be more nuanced that just considering disordering of the graphene-like network. Indeed, structural heterogeneity, interfacial coupling, and functionalization can modulate the Raman spectrum through resonance enhancement, charge transfer, strain, and sp³ bonding \cite{MADITO2025103814}.\\
For this reason, it is worth noting that the D line shift is coherent with the indication extracted from the ratio $I_D/I_G$ and thus can be related with the mechanistic interpretation we have hypothesized for the influence of precursor and electrolyte concentration on the resulting nanoparticle size. It can be inferred that the slower growth kinetics of smaller nanoparticles is reflected in a more ordered structure at the atomic scale.\\

More challenging is the discussion regarding the information that can be extracted from the signal associated with \textit{sp} carbon.\

CAWs can host a collective oscillation along the chain, often referred to as the Effective Conjugation Coordinate (ECC) mode. The ECC is Raman active, and its frequency falls in the region between 1800-2100 cm$^{-1}$. Polyynic CAWs exhibit a higher-frequency ECC (i.e. typically higher than 2100-2200 cm$^{-1}$), while cumulenic CAWs tend to present Raman signals at lower wavenumbers. In both cases, longer chains typically imply a more relaxed ECC mode \cite{casari_carbyne_2018}. Theoretical studies on the formation of cluster-assembled films suggest that \textit{sp-sp²}  carbon structures are formed by disordered \textit{sp²} domains encapsulating polyynic \textit{sp} carbon chains with different lengths \cite{Bogana_2005,PhysRevB.76.134119}. It must be noted that these results consider growth in the plasma phase and thus may be of limited validity to the present work where it would be more realistic to hypothesize that the \textit{sp} chains act as linkers between \textit{sp²} aggregates. Calculations \cite{https://doi.org/10.1002/pssb.200983946} also  suggest that the \textit{sp} band observed in amorphous film can be deconvolved as the sum of several signals originating from the coexistence of chains of different lengths and with different curvature and torsional angles. Obviously, this model is somewhat simplified, and other factors such as strain could also play a role. Due to the small amount of material achieved in this work, it was only possible to reliably discern two components ($sp^\prime$ and $sp^{\prime\prime}$) , as mentioned earlier. Nonetheless, if we assume the main contribution to the signal frequency to be from chain length, the shift in $\tilde{\nu}_{sp^\prime}$ and $\tilde{\nu}_{sp^{\prime\prime}}$ evident from Fig. \ref{fig:raman}c can be interpreted as a broadening of the length distribution of the \textit{sp} linkers in the amorphous structure. A slower reaction kinetics would then not only result in smaller nanoparticles but would allow to better preserve the chain length distribution present in the precursor solution.\
Within this framework, $I_{sp^\prime}/I_{sp^{\prime\prime}}$ can be taken as a qualitative indicator of ratio between the number of longer and shorter chains. If we assume that a larger fraction of longer chains in the amorphous matrix (i.e. larger $I_{sp^\prime}$) will be responsible for a higher overall number of \textit{sp}-carbon atoms, $I_{sp^\prime}/I_{sp^{\prime\prime}}$ should correlate to the fraction of \textit{sp}-hybridized carbon atoms $X_{sp}$. Now, by assuming as a zero-order approximation a linear relationship between $X_{sp}$ and $I_{sp^\prime}/I_{sp^{\prime\prime}}$, we can fit the data in Fig. \ref{fig:raman}d. Indeed, a correlation seems to hold between these two parameters, suggesting $I_{sp^\prime}/I_{sp^{\prime\prime}}$ as qualitative indicator of the \textit{sp} carbon content in amorphous \textit{sp-sp²} carbon materials.\

The fraction of \textit{sp}-hybridized atoms has been evaluated following the method proposed by D'Urso and co-workers for cluster-assembled films \cite{DURSO20062093}. By postulating that the structure of the material studied in this work is similar to what achieved by D'Urso and thus that the \textit{sp\textsuperscript{3}} content is negligible, Equation \ref{eq:cross} holds, where the ratio between the scattering cross sections is $\sigma_{sp}/\sigma_{sp^2} = 0.48$ for the 514 nm excitation wavelength.\

\begin{equation}
    \frac{X_{sp}}{X_{sp^2}} = \frac{I_{sp}}{I_{sp^2}} \cdot \frac{\sigma_{sp^2}}{\sigma_{sp}}\quad \text{where} \quad X_{sp} + X_{sp^2} = 1
    \label{eq:cross}
\end{equation}

By evaluating the area under the \textit{sp} and \textit{sp²} bands ($I_{sp}$ and $I_{sp^2}$), the fractions of \textit{sp} and \textit{sp²} hybridized atoms ($X_{sp}$ and $X_{sp^2}$) can be retrieved. The average values for the different experimental runs are reported in Table \ref{tab:spcont}. It must be noted that a more conservative estimation of the ratio $\sigma_{sp}/\sigma_{sp^2}$ has been proposed by Ravagnan and co-workers, even though Raman and near edge X-ray absorption fine structure spectroscopy measurements were conducted on separate samples \cite{RAVAGNAN20061518}. It would be interesting to perform synchrotron-based measurements on \textit{sp-sp²} carbon synthesized through electrochemical methods to better asses the proposed similarity with the structure of films deposited \textit{via} physical techniques. \

\begin{table}[h!]
    \centering
    \begin{tabular}{r|c c c c c}
         Diameter (nm)&  $161 \pm27$ & $125\pm21$ & $60\pm13$ & $29\pm9$ &  $23\pm3$\\
         $X_{sp}$ (\%)&  $59.6\pm0.4$ & $61.1\pm1.6$ & $54.7\pm1.7$ & $55.8\pm3.8$ & $60.9\pm1.1$    \end{tabular}
    \caption{Average \textit{sp} carbon content extracted from the Raman spectra.}
    \label{tab:spcont}
\end{table}

To further investigate the mechanism behind the formation of nanoparticles, the same electrochemical process was performed on size-selected polyynes, namely \ch{HC8H}, \ch{HC10H}, and \ch{HC12H}. Figure \ref{fig:raman}e shows the Raman spectra acquired on the electrode surface (a SEM micrograph is shown in the inset). Clearly, longer chains yield a material with a stronger \textit{sp} Raman fingerprint and result in a shift of the \textit{sp} band towards lower frequency. Following the analysis proposed in Ref. \cite{https://doi.org/10.1002/pssb.200983946}, it is not possible to directly link each of the components in the spectra to a single chain length based on the frequency of the ECC mode of H-capped polyynes. However, the range is coherent with the presence of \textit{sp} linkers between 12 and 2 atoms.  This suggests a possible interpretation: charge transfer could be responsible for the activation of the hydrogen end groups, thus favoring chain termination reactivity. This, in turn, would result in the preservation of the starting chain length in the final amorphous network. Thus, starting from a solution containing longer polyynes would result in nanoparticles with a larger fraction of \textit{sp} carbon atoms.\\ 

A final remark can be made about the stability of the \textit{sp} component in the nanoparticles. As evidenced from the Raman spectra in Fig. \ref{fig:raman}f, there is a relatively small decrease in the signal associated to the \textit{sp} component even after six months of exposure to ambient conditions. This is quite remarkable as previous examples of amorphous \textit{sp-sp²} carbon are only stable under high vacuum and for a limited time \cite{PhysRevLett.89.285506}. It is possible that the electrochemical synthesis route is able to passivate reactive sites that would otherwise induce the degradation of the \textit{sp} linkers towards more stable \textit{sp²} domains. This mechanism would also explain the apparently higher stability shown by smaller nanoparticles as they are generated by a slower kinetics, likely linked to a better passivation. While this enhancement in stability opens to new possibility to implement \textit{sp} carbon nanostructures in future devices, a more detailed characterization is needed. The exposure to aggressive chemical agents as well as high temperature in oxidizing atmosphere are all conditions that must be tested in future studies.

\section{Conclusions and Outlook}
In this work, the charge-transfer-induced reactivity of polyynic carbon atomic wires was explored. The electrochemical reduction of polydispersed H-capped polyynes in solution was demonstrated to yield amorphous carbon nanoparticles. The diameter was found to be tunable by suitably selecting the precursor and the supporting electrolyte concentration. Raman spectroscopy highlighted the significant presence of \textit{sp}-hybridized carbon atoms in the final structure.\

This phenomenon was attributed to the reductive activation of the hydrogen end groups, which is hypothesized to preserve the chain length during the cross-linking process. This hypothesis was further supported by analyses performed on size-selected polyynes. Raman spectroscopy also allowed the investigation of structural disorder in the amorphous nanoparticles, revealing that smaller nanoparticles are characterized by a more ordered \textit{sp²} network containing a broader range of \textit{sp} carbon chains.\

Unlike previous examples of amorphous \textit{sp-sp²} carbon obtained through physical methods, the electrochemical process explored in this study yielded a material that is stable in air and can be tested \textit{ex situ}, opening opportunities for future studies toward a deeper understanding of the physics underlying this class of materials with strong application potential.\

However, many aspects of this novel structure remain unclear. For instance, the mechanism behind the increased stability achieved through electrochemical methods needs to be further clarified. Additional studies are also required to investigate the effect of other process parameters, most importantly the applied potential: more negative potentials could increase the nucleation rate, further decreasing nanoparticle size. With suitable optimization, the diameter could potentially be reduced to the point of inducing quantum dot behavior. Furthermore, in this work, noble metal electrodes were used; however, the electrode material itself may play an unexplored role in polyyne reactivity.\

\section*{Supporting Information}
High-resolution SEM images and fitted Raman spectra of the samples (PDF).

\section*{Acknowledgments}
Y. Z. acknowledges Horizon Europe for the Marie Sklodowska-Curie postdoctoral Fellowship grant no. 101065920 – SCCAMC, C. S. C. acknowledges funding by the project funded under the National Recovery and Resilience Plan (NRRP), Mission 4 Component 2 Investment 1.3 Call for Tender 1561 of 11.10.2022 of Ministero dell’Università e della Ricerca (MUR), funded by the European Union NextGenerationEU Award Project Code PE0000021, Concession Decree 1561 of 11.10.2022 adopted by Ministero dell’Università e della Ricerca (MUR), CUP D43C22003090001, Project ‘‘Network 4 Energy Sustainable Transition (NEST)’’.

\clearpage
\printbibliography

@article{ravagnan_influence_2007,
	title = {Influence of {Cumulenic} {Chains} on the {Vibrational} and {Electronic} {Properties} of sp/sp2 {Amorphous} {Carbon}},
	volume = {98},
	shorttitle = {Influence of {Cumulenic} {Chains} on the {Vibrational} and {Electronic} {Properties} of {\textless}span class="aps-inline-formula"{\textgreater}{\textless}math xmlns="http},
	doi = {10.1103/PhysRevLett.98.216103},
	number = {21},
	journal = {Physical Review Letters},
	author = {Ravagnan, L.},
	year = {2007},
}

@article{park_carbyne_2013,
	title = {Carbyne bundles for a lithium-ion-battery anode},
	volume = {63},
	issn = {1976-8524},
	url = {https://doi.org/10.3938/jkps.63.1014},
	doi = {10.3938/jkps.63.1014},
	
	number = {5},
	journal = {Journal of the Korean Physical Society},
	author = {Park, M. and Lee, H.},
	month = sep,
	year = {2013},
	pages = {1014--1018},
}

@article{casari_carbyne_2018,
	title = {Carbyne: from the elusive allotrope to stable carbon atom wires},
	volume = {8},
	issn = {2159-6859, 2159-6867},
	shorttitle = {Carbyne},
	url = {https://www.cambridge.org/core/journals/mrs-communications/article/abs/carbyne-from-the-elusive-allotrope-to-stable-carbon-atom-wires/EF8BBD1E4CEAD017B6984ACD33C97FC5},
	doi = {10.1557/mrc.2018.48},
	
	language = {en},
	number = {2},
	urldate = {2025-09-24},
	journal = {MRS Communications},
	author = {Casari, C. S. and Milani, A.},
	month = jun,
	year = {2018},
	pages = {207--219},
}

@article{shi_confined_2016,
	title = {Confined linear carbon chains as a route to bulk carbyne},
	volume = {15},
	copyright = {2016 Springer Nature Limited},
	issn = {1476-4660},
	url = {https://www.nature.com/articles/nmat4617},
	doi = {10.1038/nmat4617},
	
	language = {en},
	number = {6},
	urldate = {2025-09-24},
	journal = {Nature Materials},
	author = {Shi, L. and Rohringer, P. and Suenaga, K. and Niimi, Y. and Kotakoski, J. and Meyer, J. C. and Peterlik, H. and Wanko, M. and Cahangirov, S. and Rubio, A. and Lapin, Z. J. and Novotny, L. and Ayala, P. and Pichler, T.},
	month = jun,
	year = {2016},
	note = {Publisher: Nature Publishing Group},
	pages = {634--639},
}

@article{marabotti_electron-phonon_2022,
	title = {Electron-phonon coupling and vibrational properties of size-selected linear carbon chains by resonance {Raman} scattering},
	volume = {13},
	copyright = {2022 The Author(s)},
	issn = {2041-1723},
	url = {https://www.nature.com/articles/s41467-022-32801-3},
	doi = {10.1038/s41467-022-32801-3},
	
	language = {en},
	number = {1},
	urldate = {2025-09-24},
	journal = {Nature Communications},
	author = {Marabotti, P. and Tommasini, M. and Castiglioni, C. and Serafini, P. and Peggiani, S. and Tortora, M. and Rossi, B. and Li Bassi, A. and Russo, V. and Casari, C. S.},
	month = aug,
	year = {2022},
	note = {Publisher: Nature Publishing Group},
	pages = {5052},
}

@article{eisler_polyynes_2005,
	title = {Polyynes as a {Model} for {Carbyne}: {Synthesis}, {Physical} {Properties}, and {Nonlinear} {Optical} {Response}},
	volume = {127},
	issn = {0002-7863},
	shorttitle = {Polyynes as a {Model} for {Carbyne}},
	url = {https://doi.org/10.1021/ja044526l},
	doi = {10.1021/ja044526l},
	
	number = {8},
	urldate = {2025-10-02},
	journal = {Journal of the American Chemical Society},
	author = {Eisler, S. and Slepkov, A. D. and Elliott, E. and Luu, T. and McDonald, R. and Hegmann, F. A. and Tykwinski, R. R.},
	month = mar,
	year = {2005},
	note = {Publisher: American Chemical Society},
	pages = {2666--2676},
}

@article{peggiani_situ_2020,
	title = {\textit{{In} situ} synthesis of polyynes in a polymer matrix \textit{via} pulsed laser ablation in a liquid},
	volume = {1},
	issn = {2633-5409},
	url = {https://xlink.rsc.org/?DOI=D0MA00545B},
	doi = {10.1039/D0MA00545B},
	
	language = {en},
	number = {8},
	urldate = {2025-10-06},
	journal = {Materials Advances},
	author = {Peggiani, S. and Facibeni, A. and Milani, A. and Castiglioni, C. and Russo, V. and Li Bassi, A. and Casari, C. S.},
	year = {2020},
	pages = {2729--2736},
}

@article{wang_carbon_2020,
	title = {Carbon nanomaterials with sp or/and sp hybridization in energy conversion and storage applications: {A} review},
	volume = {26},
	issn = {24058297},
	shorttitle = {Carbon nanomaterials with sp or/and sp hybridization in energy conversion and storage applications},
	url = {https://linkinghub.elsevier.com/retrieve/pii/S2405829719310414},
	doi = {10.1016/j.ensm.2019.11.006},
	
	language = {en},
	urldate = {2025-11-05},
	journal = {Energy Storage Materials},
	author = {Wang, Y. and Yang, P. and Zheng, L. and Shi, X. and Zheng, H.},
	month = apr,
	year = {2020},
	pages = {349--370},
}

@article{CATALDO200782,
title = {Kinetics of polyynes formation with the submerged carbon arc},
journal = {Journal of Electroanalytical Chemistry},
volume = {602},
number = {1},
pages = {82-90},
year = {2007},
issn = {1572-6657},
doi = {https://doi.org/10.1016/j.jelechem.2006.12.005},
url = {https://www.sciencedirect.com/science/article/pii/S0022072806006802},
author = {F. Cataldo and O. Ursini and G. Angelini},

}

@article{EASTMOND19724601,
title = {Silylation as a protective method for terminal alkynes in oxidative couplings: A general synthesis of the parent polyynes H(C C)nH (n = 4–10, 12)},
journal = {Tetrahedron},
volume = {28},
number = {17},
pages = {4601-4616},
year = {1972},
issn = {0040-4020},
doi = {https://doi.org/10.1016/0040-4020(72)80041-3},
url = {https://www.sciencedirect.com/science/article/pii/0040402072800413},
author = {R. Eastmond and T.R. Johnson and D.R.M. Walton},

}

@article{doi:10.1021/acs.jpcc.6b03071,
author = {Agarwal, N. R. and Lucotti, A. and Tommasini, M. and Chalifoux, W. A. and Tykwinski, R. R.},
title = {Nonlinear Optical Properties of Polyynes: An Experimental Prediction for Carbyne},
journal = {The Journal of Physical Chemistry C},
volume = {120},
number = {20},
pages = {11131-11139},
year = {2016},
doi = {10.1021/acs.jpcc.6b03071},

URL = { 
    
        https://doi.org/10.1021/acs.jpcc.6b03071
    
    

},
eprint = { 
    
        https://doi.org/10.1021/acs.jpcc.6b03071
    
    

}

}

@article{MARABOTTI2024118503,
title = {Synchrotron-based UV resonance Raman spectroscopy probes size confinement, termination effects, and anharmonicity of carbon atomic wires},
journal = {Carbon},
volume = {216},
pages = {118503},
year = {2024},
issn = {0008-6223},
doi = {https://doi.org/10.1016/j.carbon.2023.118503},
url = {https://www.sciencedirect.com/science/article/pii/S0008622323007480},
author = {P. Marabotti and M. Tommasini and C. Castiglioni and S. Peggiani and P. Serafini and B. Rossi and A. {Li Bassi} and V. Russo and C.S. Casari},

}

@article{RAVAGNAN20061518,
title = {Quantitative evaluation of sp/sp2 hybridization ratio in cluster-assembled carbon films by in situ near edge X-ray absorption fine structure spectroscopy},
journal = {Carbon},
volume = {44},
number = {8},
pages = {1518-1524},
year = {2006},
issn = {0008-6223},
doi = {https://doi.org/10.1016/j.carbon.2005.12.015},
url = {https://www.sciencedirect.com/science/article/pii/S0008622305007384},
author = {L. Ravagnan and G. Bongiorno and D. Bandiera and E. Salis and P. Piseri and P. Milani and C. Lenardi and M. Coreno and M. {de Simone} and K.C. Prince},

}

@article{doi:10.1021/acs.nanolett.0c02632,
author = {Tschannen, C. D. and Gordeev, G. and Reich, S. and Shi, L. and Pichler, T. and Frimmer, M. and Novotny, L. and Heeg, S.},
title = {Raman Scattering Cross Section of Confined Carbyne},
journal = {Nano Letters},
volume = {20},
number = {9},
pages = {6750-6755},
year = {2020},
doi = {10.1021/acs.nanolett.0c02632},
    note ={PMID: 32786933},

URL = { 
    
        https://doi.org/10.1021/acs.nanolett.0c02632
    
    

},
eprint = { 
    
        https://doi.org/10.1021/acs.nanolett.0c02632
    
    

}

}

@article{doi:10.1021/acsnano.1c03893,
author = {Tschannen, C. D. and Frimmer, M. and Gordeev, G. and Vasconcelos, T. L. and Shi, L. and Pichler, T. and Reich, S. and Heeg, S. and Novotny, L.},
title = {Anti-Stokes Raman Scattering of Single Carbyne Chains},
journal = {ACS Nano},
volume = {15},
number = {7},
pages = {12249-12255},
year = {2021},
doi = {10.1021/acsnano.1c03893},
    note ={PMID: 34254777},

URL = { 
    
        https://doi.org/10.1021/acsnano.1c03893
    
    

},
eprint = { 
    
        https://doi.org/10.1021/acsnano.1c03893
    
    

}

}

@article{hu_supermultiplexed_2018,
	title = {Supermultiplexed optical imaging and barcoding with engineered polyynes},
	volume = {15},
	copyright = {2018 Springer Nature America, Inc.},
	issn = {1548-7105},
	url = {https://www.nature.com/articles/nmeth.4578},
	doi = {10.1038/nmeth.4578},
	language = {en},
	number = {3},
	urldate = {2025-11-06},
	journal = {Nature Methods},
	author = {Hu, F. and Zeng, C. and Long, R. and Miao, Y. and Wei, L. and Xu, Q. and Min, W.},
	month = mar,
	year = {2018},
	note = {Publisher: Nature Publishing Group},
	pages = {194--200},
}

@article{doi:10.1021/acscentsci.3c01090,
author = {Arora, A. and Baksi, S. D. and Weisbach, N. and Amini, H. and Bhuvanesh, N. and Gladysz, J. A.},
title = {Monodisperse Molecular Models for the sp Carbon Allotrope Carbyne; Syntheses, Structures, and Properties of Diplatinum Polyynediyl Complexes with PtC20Pt to PtC52Pt Linkages},
journal = {ACS Central Science},
volume = {9},
number = {12},
pages = {2225-2240},
year = {2023},
doi = {10.1021/acscentsci.3c01090},

URL = { 
    
        https://doi.org/10.1021/acscentsci.3c01090
    
    

},
eprint = { 
    
        https://doi.org/10.1021/acscentsci.3c01090
    
    

}

}

@article{doi:10.1021/acs.jpcc.5c02960,
author = {Melesi, S. and Pińkowski, P. and Pigulski, B. and Gulia, N. and Szafert, S. and Bertarelli, C. and Castiglioni, C. and Casari, C. S.},
title = {Probing the Stability of Halogenated Carbon Atomic Wires in Electrospun Nanofibers via Raman Spectroscopy},
journal = {The Journal of Physical Chemistry C},
volume = {129},
number = {28},
pages = {12916-12926},
year = {2025},
doi = {10.1021/acs.jpcc.5c02960},

URL = { 
    
        https://doi.org/10.1021/acs.jpcc.5c02960
    
    

},
eprint = { 
    
        https://doi.org/10.1021/acs.jpcc.5c02960
    
    

}

}

@article{ghosh_cumulenic_2025,
	title = {Cumulenic sp‐{Carbon} atomic wires wrapped with polymer for supercapacitor application},
	volume = {234},
	issn = {0008-6223},
	url = {https://www.sciencedirect.com/science/article/pii/S0008622324011710},
	doi = {https://doi.org/10.1016/j.carbon.2024.119952},
	
	journal = {Carbon},
	author = {Ghosh, S. and Righi, M. and Melesi, S. and Qiu, Y. and Tykwinski, R. R. and Casari, C. S.},
	year = {2025},
	pages = {119952},
}

@article{scaccabarozzi_field-effect_2020,
	title = {A {Field}-{Effect} {Transistor} {Based} on {Cumulenic} sp-{Carbon} {Atomic} {Wires}},
	volume = {11},
	url = {https://doi.org/10.1021/acs.jpclett.0c00141},
	doi = {10.1021/acs.jpclett.0c00141},
	
	number = {5},
	urldate = {2025-09-22},
	journal = {The Journal of Physical Chemistry Letters},
	author = {Scaccabarozzi, A. D. and Milani, A. and Peggiani, S. and Pecorario, S. and Sun, B. and Tykwinski, R. R. and Caironi, M. and Casari, C. S.},
	month = mar,
	year = {2020},
	note = {Publisher: American Chemical Society},
	pages = {1970--1974},
}

@article{rabia_structural_2020,
	title = {Structural, {Electronic}, and {Vibrational} {Properties} of a {Two}-{Dimensional} {Graphdiyne}-like {Carbon} {Nanonetwork} {Synthesized} on {Au}(111): {Implications} for the {Engineering} of sp-sp2 {Carbon} {Nanostructures}},
	volume = {3},
	shorttitle = {Structural, {Electronic}, and {Vibrational} {Properties} of a {Two}-{Dimensional} {Graphdiyne}-like {Carbon} {Nanonetwork} {Synthesized} on {Au}(111)},
	url = {https://doi.org/10.1021/acsanm.0c02665},
	doi = {10.1021/acsanm.0c02665},
	
	number = {12},
	urldate = {2025-09-22},
	journal = {ACS Appl. Nano Mater.},
	author = {Rabia, A. and Tumino, F. and Milani, A. and Russo, V. and Li Bassi, A. and Bassi, N. and Lucotti, A. and Achilli, S. and Fratesi, G. and Manini, N. and Onida, G. and Sun, Q. and Xu, W. and Casari, C. S.},
	month = dec,
	year = {2020},
	note = {Publisher: American Chemical Society},
	pages = {12178--12187},
}

@article{pan_carbyne_2015,
	title = {Carbyne with finite length: {The} one-dimensional \textit{sp} carbon},
	volume = {1},
	issn = {2375-2548},
	shorttitle = {Carbyne with finite length},
	url = {https://www.science.org/doi/10.1126/sciadv.1500857},
	doi = {10.1126/sciadv.1500857},
	
	language = {en},
	number = {9},
	urldate = {2025-11-07},
	journal = {Science Advances},
	author = {Pan, B. and Xiao, J. and Li, J. and Liu, P. and Wang, C. and Yang, G.},
	month = oct,
	year = {2015},
	pages = {e1500857},
}

@article{YANG2022100692,
title = {Synthesis, properties, and applications of carbyne nanocrystals},
journal = {Materials Science and Engineering: R: Reports},
volume = {151},
pages = {100692},
year = {2022},
issn = {0927-796X},
doi = {https://doi.org/10.1016/j.mser.2022.100692},
url = {https://www.sciencedirect.com/science/article/pii/S0927796X22000316},
author = {G. Yang},

}

@article{PhysRevLett.89.285506,
  title = {Cluster-Beam Deposition and in situ Characterization of Carbyne-Rich Carbon Films},
  author = {Ravagnan, L. and Siviero, F. and Lenardi, C. and Piseri, P. and Barborini, E. and Milani, P. and Casari, C. S. and Li Bassi, A. and Bottani, C. E.},
  journal = {Phys. Rev. Lett.},
  volume = {89},
  issue = {28},
  pages = {285506},
  numpages = {4},
  year = {2002},
  month = {Dec},
  publisher = {American Physical Society},
  doi = {10.1103/PhysRevLett.89.285506},
  url = {https://link.aps.org/doi/10.1103/PhysRevLett.89.285506}
}

@article{casari_chemical_2004,
	title = {Chemical and thermal stability of carbyne-like structures in cluster-assembled carbon films},
	volume = {69},
	copyright = {http://link.aps.org/licenses/aps-default-license},
	issn = {1098-0121, 1550-235X},
	url = {https://link.aps.org/doi/10.1103/PhysRevB.69.075422},
	doi = {10.1103/PhysRevB.69.075422},
	language = {en},
	number = {7},
	urldate = {2025-11-03},
	journal = {Physical Review B},
	author = {Casari, C. S. and Li Bassi, A. and Ravagnan, L. and Siviero, F. and Lenardi, C. and Piseri, P. and Bongiorno, G. and Bottani, C. E. and Milani, P.},
	month = feb,
	year = {2004},
	pages = {075422},
}

@article{Bettini_2016,
doi = {10.1088/0957-4484/27/11/115403},
url = {https://doi.org/10.1088/0957-4484/27/11/115403},
year = {2016},
month = {feb},
publisher = {IOP Publishing},
volume = {27},
number = {11},
pages = {115403},
author = {Bettini, L. G. and Della Foglia, F. and Piseri, P. and Milani, P.},
title = {Interfacial properties of a carbyne-rich nanostructured carbon thin film in ionic liquid},
journal = {Nanotechnology},
}

@article{COMPAGNINI20122362,
title = {Deposition of few layer graphene nanowalls at the electrodes during electric field-assisted laser ablation of carbon in water},
journal = {Carbon},
volume = {50},
number = {6},
pages = {2362-2365},
year = {2012},
issn = {0008-6223},
doi = {https://doi.org/10.1016/j.carbon.2012.01.038},
url = {https://www.sciencedirect.com/science/article/pii/S0008622312000693},
author = {G. Compagnini and M. Sinatra and P. Russo and G. C. Messina and O. Puglisi and S. Scalese},
}

@article{RUSSO201754,
title = {Carbon nanowalls: A new material for resistive switching memory devices},
journal = {Carbon},
volume = {120},
pages = {54-62},
year = {2017},
issn = {0008-6223},
doi = {https://doi.org/10.1016/j.carbon.2017.05.004},
url = {https://www.sciencedirect.com/science/article/pii/S0008622317304499},
author = {P. Russo and M. Xiao and N. Y. Zhou},

}

@article{PhysRevB.61.14095,
  title = {Interpretation of Raman spectra of disordered and amorphous carbon},
  author = {Ferrari, A. C. and Robertson, J.},
  journal = {Phys. Rev. B},
  volume = {61},
  issue = {20},
  pages = {14095--14107},
  numpages = {0},
  year = {2000},
  month = {May},
  publisher = {American Physical Society},
  doi = {10.1103/PhysRevB.61.14095},
  url = {https://link.aps.org/doi/10.1103/PhysRevB.61.14095}
}

@article{kastner_reductive_1995,
	title = {Reductive {Preparation} of {Carbyne} with {High} {Yield}. {An} in {Situ} {Raman} {Scattering} {Study}},
	volume = {28},
	issn = {0024-9297, 1520-5835},
	url = {https://pubs.acs.org/doi/abs/10.1021/ma00105a048},
	doi = {10.1021/ma00105a048},
	language = {en},
	number = {1},
	urldate = {2025-11-07},
	journal = {Macromolecules},
	author = {Kastner, J. and Kuzmany, H. and Kavan, L. and Dousek, F. P. and Kuerti, J.},
	month = jan,
	year = {1995},
	pages = {344--353},
}

@article{kijima_novel_1996,
	title = {Novel approach for synthesis of a carbyne film by electrochemical reduction of hexachlorobuta-1,3-diene},
	issn = {1359-7345, 1364-548X},
	url = {https://xlink.rsc.org/?DOI=cc9960002273},
	doi = {10.1039/cc9960002273},
	language = {en},
	number = {19},
	urldate = {2025-11-07},
	journal = {Chemical Communications},
	author = {Kijima, M. and Toyabe, T. and Shirakawa, H.},
	year = {1996},
	pages = {2273},
}

@article{hlavaty_modification_1997,
	title = {Modification of electrochemical carbon by \textit{in situ} generated carbenes},
	volume = {35},
	issn = {0008-6223},
	url = {https://www.sciencedirect.com/science/article/pii/S0008622397811190},
	doi = {10.1016/S0008-6223(97)81119-0},
	
	number = {1},
	urldate = {2025-11-07},
	journal = {Carbon},
	author = {Hlavatý, J. and Kavan, L.},
	month = jan,
	year = {1997},
	pages = {127--131},
}

@article{CATALDO20052792,
title = {A method for synthesizing polyynes in solution},
journal = {Carbon},
volume = {43},
number = {13},
pages = {2792-2800},
year = {2005},
issn = {0008-6223},
doi = {https://doi.org/10.1016/j.carbon.2005.05.024},
url = {https://www.sciencedirect.com/science/article/pii/S0008622305003118},
author = {F. Cataldo},
}

@article{MARABOTTI2022219,
title = {In situ surface-enhanced Raman spectroscopy to investigate polyyne formation during pulsed laser ablation in liquid},
journal = {Carbon},
volume = {189},
pages = {219-229},
year = {2022},
issn = {0008-6223},
doi = {https://doi.org/10.1016/j.carbon.2021.12.060},
url = {https://www.sciencedirect.com/science/article/pii/S0008622321012197},
author = {P. Marabotti and S. Peggiani and A. Facibeni and P. Serafini and A. Milani and V. Russo and A. {Li Bassi} and C.S. Casari}
}

@article{DURSO20062093,
title = {sp/sp2 bonding ratio in sp rich amorphous carbon thin films},
journal = {Carbon},
volume = {44},
number = {10},
pages = {2093-2096},
year = {2006},
issn = {0008-6223},
doi = {https://doi.org/10.1016/j.carbon.2006.04.016},
url = {https://www.sciencedirect.com/science/article/pii/S0008622306002119},
author = {L. D’Urso and G. Compagnini and O. Puglisi}
}

@article{PhysRevB.76.134119,
  title = {Growth of $\mathrm{sp}\text{\ensuremath{-}}{\mathrm{sp}}^{2}$ nanostructures in a carbon plasma},
  author = {Yamaguchi, Y. and Colombo, L. and Piseri, P. and Ravagnan, L. and Milani, P.},
  journal = {Phys. Rev. B},
  volume = {76},
  issue = {13},
  pages = {134119},
  numpages = {7},
  year = {2007},
  month = {Oct},
  publisher = {American Physical Society},
  doi = {10.1103/PhysRevB.76.134119},
  url = {https://link.aps.org/doi/10.1103/PhysRevB.76.134119}
}

@article{Bogana_2005,
doi = {10.1088/1367-2630/7/1/081},
url = {https://doi.org/10.1088/1367-2630/7/1/081},
year = {2005},
month = {mar},
publisher = {},
volume = {7},
number = {1},
pages = {81},
author = {Bogana, M. and Ravagnan, L. and Casari, C. S. and Zivelonghi, A. and Baserga, A. and Li Bassi, A. and Bottani, C. E. and Vinati, S. and Salis, E. and Piseri, P. and Barborini, E. and Colombo, L. and Milani, P.},
title = {Leaving the fullerene road: presence and stability of sp chains in sp2 carbon clusters and cluster-assembled solids},
journal = {New Journal of Physics},

}

@article{https://doi.org/10.1002/pssb.200983946,
author = {Onida, G. and Manini, N. and Ravagnan, L. and Cinquanta, E. and Sangalli, D. and Milani, P.},
title = {Vibrational properties of sp carbon atomic wires in cluster-assembled carbon films},
journal = {physica status solidi (b)},
volume = {247},
number = {8},
pages = {2017-2021},
doi = {https://doi.org/10.1002/pssb.200983946},
url = {https://onlinelibrary.wiley.com/doi/abs/10.1002/pssb.200983946},
eprint = {https://onlinelibrary.wiley.com/doi/pdf/10.1002/pssb.200983946},

year = {2010}
}

@Article{B922733D,
author ="Li, G. and Li, Y. and Liu, H. and Guo, Y. and Li, Y. and Zhu, D.",
title  ="Architecture of graphdiyne nanoscale films",
journal  ="Chem. Commun.",
year  ="2010",
volume  ="46",
issue  ="19",
pages  ="3256-3258",
publisher  ="The Royal Society of Chemistry",
doi  ="10.1039/B922733D",
url  ="http://dx.doi.org/10.1039/B922733D",
}

@article{doi:10.1021/acs.jpcc.1c04238,
author = {Serafini, P. and Milani, A. and Proserpio, D. M. and Casari, C. S.},
title = {Designing All Graphdiyne Materials as Graphene Derivatives: Topologically Driven Modulation of Electronic Properties},
journal = {The Journal of Physical Chemistry C},
volume = {125},
number = {33},
pages = {18456-18466},
year = {2021},
doi = {10.1021/acs.jpcc.1c04238},

URL = { 
    
        https://doi.org/10.1021/acs.jpcc.1c04238
    
    

},
eprint = { 
    
        https://doi.org/10.1021/acs.jpcc.1c04238
    
    

}

}

@article{KIJIMA19951837,
title = {Electrochemical synthesis of carbyne catalyzed by nickel complex},
journal = {Synthetic Metals},
volume = {71},
number = {1},
pages = {1837-1840},
year = {1995},
note = {Proceedings of the International Conference on Science and Technology of Synthetic Metals (ICSM '94)},
issn = {0379-6779},
doi = {https://doi.org/10.1016/0379-6779(94)03074-G},
url = {https://www.sciencedirect.com/science/article/pii/037967799403074G},
author = {M. Kijima and Y. Sakai and H. Shirakawa}
}

@article{JAYSWAL2023337,
title = {Synthesis of fluorescent carbon nanoparticles by dispersion polymerization of acetylene},
journal = {Nanoscale Advances},
volume = {5},
number = {2},
pages = {337-343},
year = {2023},
issn = {2516-0230},
doi = {https://doi.org/10.1039/d2na00619g},
url = {https://www.sciencedirect.com/science/article/pii/S2516023023021792},
author = {V. K. Jayswal and A. M. Ritcey and J. Morin}
}

@article{10.1063/1.362745,
    author = {Schwan, J. and Ulrich, S. and Batori, V. and Ehrhardt, H. and Silva, S. R. P.},
    title = {Raman spectroscopy on amorphous carbon films},
    journal = {Journal of Applied Physics},
    volume = {80},
    number = {1},
    pages = {440-447},
    year = {1996},
    month = {07},
    issn = {0021-8979},
    doi = {10.1063/1.362745},
    url = {https://doi.org/10.1063/1.362745},
    eprint = {https://pubs.aip.org/aip/jap/article-pdf/80/1/440/18685905/440_1_online.pdf},
}

@article{LI2013328,
title = {Anomalous solubility of oxygen in acetonitrile/water mixture containing tetra-n-butylammonium perchlorate supporting electrolyte; the solubility and diffusion coefficient of oxygen in anhydrous acetonitrile and aqueous mixtures},
journal = {Journal of Electroanalytical Chemistry},
volume = {688},
pages = {328-335},
year = {2013},
note = {Special Issue in Honor of Professors Chuansin Cha and Zhaowu Tian},
issn = {1572-6657},
doi = {https://doi.org/10.1016/j.jelechem.2012.07.039},
url = {https://www.sciencedirect.com/science/article/pii/S1572665712003001},
author = {Q. Li and C. Batchelor-McAuley and N. S. Lawrence and R. S. Hartshorne and R. G. Compton},

}

@article{MADITO2025103814,
title = {Revisiting the Raman disorder band in graphene-based materials: A critical review},
journal = {Vibrational Spectroscopy},
volume = {139},
pages = {103814},
year = {2025},
issn = {0924-2031},
doi = {https://doi.org/10.1016/j.vibspec.2025.103814},
url = {https://www.sciencedirect.com/science/article/pii/S0924203125000487},
author = {M.J. Madito},

}
\clearpage

\begin{figure}[h]
    \raggedright
    \includegraphics[width=3.25in,height=1.75in]{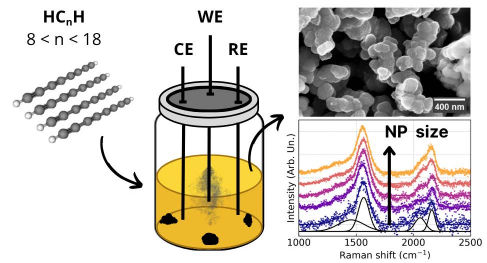}
    \captionsetup{justification=raggedright,singlelinecheck=false}
    \caption*{TOC Graphic}
\end{figure}

\includepdf[pages=-]{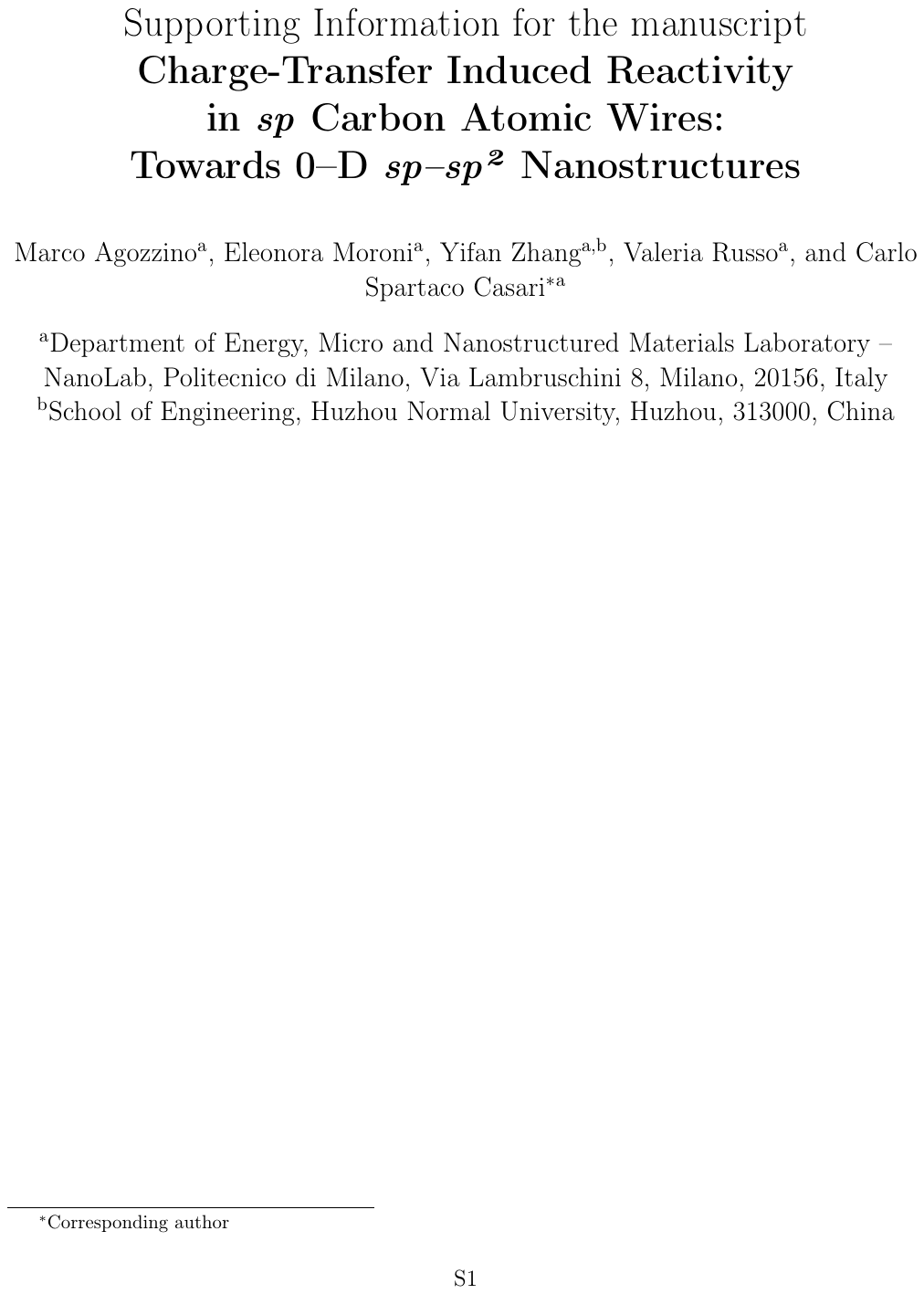}

\end{document}